\documentclass[epj,draft]{svjour}
\usepackage{epsf}
\usepackage{amsmath,amssymb}

\newcommand{\ve}[1]{\boldsymbol{#1}}

\begin{document}

\title{Interstitial Fe-Cr alloys: Tuning of magnetism by nanoscale structural control 
and by implantation of nonmagnetic atoms}
\subtitle{Interstitial Fe-Cr alloys}
\author{N.~Pavlenko\inst{1}, N.~Shcherbovskikh\inst{2}, and Z.A.~Duriagina\inst{2}}
\institute{Institute for Condensed Matter Physics, National Academy of Sciences of Ukraine, 
Svientsitsky str.1, 79011 Lviv, Ukraine;
\email{pavlenko@mailaps.org}  \and
Institute for Applied Mathematics and Fundamental Sciences,
Lviv Technical University, Ustyianowycha str.~10, 79013 Lviv, Ukraine
}


\abstract{Using the density functional theory, we perform a full atomic relaxation of the bulk
ferrite with $12.5\%$-concentration of monoatomic interstitial Cr
periodically located at the edges of the bcc Fe$_\alpha$ cell.
We show that structural relaxation in such artificially engineered 
alloys leads to significant atomic displacements and
results in the formation of novel highly stable configurations with parallel
chains of octahedrically arranged Fe.
The enhanced magnetic polarization in the low-symmetry metallic state
of this type of alloys can be externally
controlled by additional inclusion of nonmagnetic impurities like nitrogen.
We discuss possible applications of generated interstitial alloys 
in spintronic devices and propose to consider them as a basis of novel durable
types of stainless steels.}

\maketitle




\section{Introduction}

Last years demonstrate increased activities in the search for novel materials exhibiting controlled modification of electronic 
properties by inclusion or implantation of different atoms or ionic groups. A prominent example of the implantation-altered 
systems is the stainless steel. In the steels, the implantation 
of chromium, molibdenium, nitrogen and other chemical elements substantially changes the microstructure of subsurface layers and 
modify their corrosion resistance and hardness \cite{steels}.

In the development of novel efficient multifunctional materials for technological applications in the long-term devices, 
the properties like hardness, corrosion, heat resistance and other types of mechanical and chemical durability are of 
central interest \cite{mai,yokokawa}. It frequently appears in science and technology that well known materials doped
by different chemical elements exhibit unexpected physical properties not revealed previously.

As an example of such a new unexpected behavior, in the present work we consider an alloy Fe-Cr. The alloys of Fe and Cr,
doped by C, Ni and by other elements, are widely used as basic components for ferritic and martensitic steels.
Substitutional alloys of Fe and Cr have attracted 
much attention of theory and experiment due to their rich magnetic properties characterized by local antiferromagnetism 
in the proximity of Cr atoms implanted into ferromagnetic iron \cite{victora,paxton,paduani,davies}. Due to small differences 
between the atomic radii of iron and chromium, the modification of the substitutional alloy properties is
limited to the local magnetic transformation due to local changes in the electronic orbital
occupancies, without significant structural modifications. 
In contrast to the substitutional structural configurations, the interstitial Fe-Cr 
alloys considered in the present 
work contain Cr impurities which are located in the interstitial positions of the bcc lattice of Fe$_\alpha$. 
In the recent theoretical studies of the Cr intersitials in Fe-Cr alloys, different types 
of interstitial configurations were analyzed. Among them, a pair configuration $\langle 111 \rangle$  dumbbell 
is considered as the most energetically favourable which requires about 4.2eV for its formation 
under irradiation \cite{klaver,olsson}. 

In the present work, we consider a novel monoatomic interstitial configuration which contains single Cr atoms 
positioned in the centers of the edges of the bcc ferrite. In contrast to the substitutional alloys, the significant forces
due to the interstitial atoms induce substantial structural optimization which enhances the volume due 
to modified lattice constants and leads to the relaxation of the atomic positions in the unit cell. 
We find that the relaxation
of the initial bcc unit cell results in significant atomic distortions and in the formation of atomic 
chain-like structures.   
As appears in the density-functional-theory (DFT) calculations of the optimized structures, the energy gain achieved due 
to the structural relaxation of the considered interstitial 
alloy can approach 6.17~eV which makes this type of systems highly stable and 
durable. In the present work, we propose
to consider these artificially generated alloys as candidates for novel types of stainless steels.  

The fundamental difference between the industrial alloys and the alloys studied in the present work 
is the ordered and periodic character of the latters. In the industrial steels, the amorphic character of the systems is 
related to the random distribution of the impurities.
The hardening of the steels proceeds through the surface treatment and is accompanied 
by formation of granular microstructure with the spatially inhomogeneous impurity concentration and modified subsurface 
properties \cite{afm}. In the studies of the subsurface Cr-doped alloyed ferrite, we consider the supercells
containing periodically located Cr atoms in the cubic lattice 
of Fe$_\alpha$. The interstitial Cr induces significant atomic reconstruction with consequent
break of initial cubic symmetry 
and stabilization of a new lower-symmetry state. 
The appearing structural transformation has a character of a phase transition which occurs due to nanoscale tailoring of cubic 
Fe by interstitial 
inclusion of Cr atoms, the effect which can be experimentally verified by the means of modern methods like AFM spectroscopy .
 
Using the DFT-based structural optimization, we obtain
the optimized atomic microstructure of a chain-like character where the chains 
of octahedrically arranged Fe atoms are formed along the (001)-axis. We find that the competing ferromagnetic and 
antiferromagnetic interactions lead to spatially inhomogeneous spin polarization. 
The magnetization of the structurally relaxed system 
is significantly enhanced as compared to the pure ferrite without Cr inclusions. 
The obtained enhancement makes the generated alloys perspective candidates 
for spin polarizers in  spintronic applications. In the generated chain-like structures, 
the relaxation is accompanied by the formation of 
spatial channels with extremely low carrier density. 
We suggest that these channels can be considered as paths for the low-barrier-migration 
of light impurities like H, N, Li or C. 
As an example of a light atom in the interstitial alloy, 
we study of the migration paths of nonmagnetic nitrogen and calculate the energy barriers along 
the migration paths. 
We obtain a strong influence of the nonmagnetic N on the alloy magnetization. Our findings show that 
the structural modifications
due to possible nanoscale tuning of Cr impurities on the edges of bcc cubic cells of iron can 
play a central role in the control of their electronic properties.

\section{Structural relaxation of interstitial alloy Fe$_\alpha$-Cr}

\begin{figure}[th]
\epsfxsize=6.0cm {\epsffile{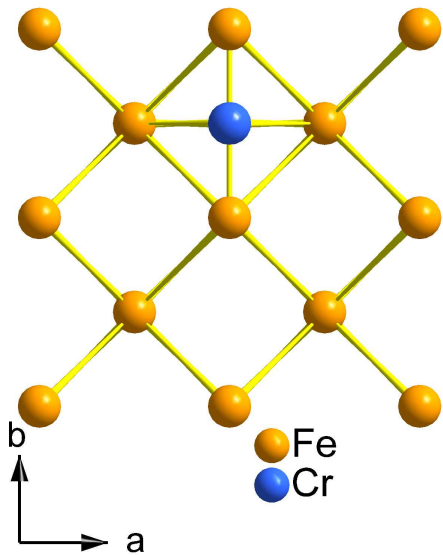}}
\caption{Schematic view of unrelaxed 2$\times$2 Fe$_\alpha$ cell which contain 12.5$\%$ of edge-centered 
interstitial Cr.  
} \label{fig1}
\end{figure}

The present studies of the electronic properties of the considered interstitial Fe-Cr alloy are based 
on the DFT calculations of the electronic structure of the systems generated by periodic translation
of specially chosen supercells. The initial supercell shown in Fig.~\ref{fig1} contains the doubled $2 \times 2$ 
cubic bcc cell of ferrite (Fe$_\alpha$) and a single Cr atom centered in one of the edges of the 
Fe$_\alpha$ cubic unit cell with the lattice constant $a=3.85$~\AA.
The obtained structure is described by a chemical formula Fe$_8$Cr 
and determines an interstitial Fe-Cr alloy with the Cr concentration 
$n=0.125$ which is typical for stainless steels. 
The presence of interstitial Cr leads to significant local forces acting on the neighbouring Fe atoms. To minimize the forces, 
the coordinates of all atoms have been relaxed.
In the present studies, the optimization of the supercell has been performed by employing the DFT approach
implemented withing the
linearized augmented plane wave (LAPW) scheme in the full potential Wien2k code \cite{wien2k}. 
To study the role of the spin polarization in the structural relaxation, 
two different relaxation procedures have been employed. In the first procedure, the atomic optimal positions are calculated 
in the local density approximation (LDA) on a $2\times 2 \times 5$ k-points grid. To explore the role of spin degrees of 
freedom in the relaxation, in the second procedure the local spin density approximation (LSDA) has been used in the
optimization of the structure.
The results of both methods of the structural relaxation are presented in Fig.~\ref{fig2}.

A central common feature which characterizes both (LDA- and LSDA-relaxed) structures is the 
clusterization of the sublattice of the iron atoms. 
In the LDA-optimized structure (Fig.~\ref{fig2}(a)), the relaxation results in
formation of a high-symmetry clusterized network. This network consists of the Fe$_6$-octahedra which form 
the square plaquettes in the $(x, y)$ ($(a,b)$) plane with Cr atoms located in the center of each plaquette.
The distance from the centered Cr to each nearest iron octaherda amounts $1.9$~\AA. 
Despite the significant displacements of the iron atoms from their initial positions, 
the net electric polarization of the cell is zero due to high structural symmetry 
C4/m obtained after the relaxation. 

The formation energy of the relaxed Fe$_8$Cr-configuration can be expressed as 
\begin{eqnarray*}
E_f({\rm LDA})=E_{{\rm tot}}({\rm Fe}_8{\rm Cr})-
8E_{{\rm tot}}({\rm Fe})-E_{{\rm tot}}({\rm Cr}),
\end{eqnarray*} 
where the last two terms identify the total energies of the bulk bcc Fe$_{\alpha}$ and Cr, respectively. 
To determine $E_{{\rm tot}}({\rm Fe})$, we have calculated the total energy 
value of the bulk Fe$_{\alpha}$ in the ferromagnetic state. As the LSDA-calculation of the spin-polarized
configurations of the bulk Cr are converged to the paramagnetic state, we consider the total energy
$E_{{\rm tot}}({\rm Cr})$ for the paramagnetic Cr.  
With these values, we find that $E_f({\rm LDA})=4.82$eV. 
To analyze the role of the relaxation, we have also calculated the energy $E_f({\rm unrel})$ of the 
formation of initial unrelaxed configuration which is equal to 5.02eV. 
As a consequence, the significant energy gain due 
to the structural relaxation 
\begin{eqnarray*}
\Delta E({\rm LDA})=E_f({\rm unrel})-E_f({\rm LDA})=0.196eV, 
\end{eqnarray*}
shows a central importance of the atomic displacements for the stability of the considered systems.

The optimization procedure based on the LSDA approach accounts for additional corrections 
due to spin polarization
and produces new ordered structural patterns presented in Fig.~\ref{fig2}(b) and Fig.~\ref{fig2}(c) 
for two different (unrelaxed $a=b=2.86$\AA\, and relaxed $a=b=3$\AA) lattice constants. 
The volume-optimized structure (c) is signified by the 13\%th increase of the unit cell 
volume due to the insertion of the interstitial Cr. 
The LSDA-optimized structural pattern is characterized by the 
chains of atomic Fe-groups along the $x$($a$)-direction, each group containing six Fe-atoms. 
The nearest chains are separated by a distance about $4$\AA\, and are connected to 
each other by the Fe-Cr bonds of the length about $2.4$\AA\, for the 
structure (b) with $a=2.86$\AA, and $2.7$\AA\, for the structure (c) with the optimized $a=3.0$\AA. 
The local antiferromagnetic ordering in the vicinity of Cr is characterized by the magnetic moments 
$\mu_{Cr}=-0.72$~$\mu_B$ 
and $\mu_{5}=2.4$~$\mu_B$ and $\mu_{6}=1.25$~$\mu_B$ of the neighbouring atoms Fe5 and Fe6, respectively. 
The magnetic moments of more distant iron atoms have the values around 2.5~$\mu_B$, which 
is close to results obtained for substitutional alloys and in pure Fe$_\alpha$ \cite{klaver}.

As compared to the tetragonal structure of the LDA-optimized system, the chain-like structure of the LSDA-relaxed supercell is 
characterized by substantially lower crystal symmetry and by the absence of the inversion center. 
In contrast to the LDA-based configurations, the formation energy of the LSDA-relaxed Fe$_8$Cr configuration 
$E_f({\rm LSDA})=E_{{\rm tot}}({\rm Fe}_8{\rm Cr})-8E_{{\rm tot}}({\rm Fe})-E_{{\rm tot}}(Cr)=-1.15$eV is negative which 
implies its high stability.
We can also calculate 
the energy gain due to the structural relaxation by the LSDA approach 
\begin{eqnarray*}
\Delta E({\rm LSDA})=E_f({\rm unrel})-E_f({\rm LSDA})=6.17eV,  
\end{eqnarray*}
which also demonstrates the high stability of the relaxed spin-polarized structure and a necessity
to account for a spin polarization in the structural optimization of the systems with strong 
magnetoelastic effect.

\subsection{Magnetic properties}

In the considered systems, we have also analyzed modification of the local magnetic properties 
due to the relaxation of the interatomic distances. 
To see how the atomic displacements influence the spin polarization of the surrounding atoms, in Fig.~\ref{fig3} we 
present the dependences of the local moments 
of Cr and of two nearest neighboring Fe on the Cr displacement along the bond [Fe5-Cr-Fe6] 
\begin{eqnarray*}
\Delta=[x({\rm Cr})-x({\rm Fe5})]-[x({\rm Cr})-x({\rm Fe5})]^0, 
\end{eqnarray*}
where $[x({\rm Cr})-x({\rm Fe5})]^0$ is the optimized
[Fe5-Cr]-bond length. 
The increase of $\Delta$ leads to the change of $\mu_{\rm Cr}$ from -0.7~$\mu_{B}$ to the value 
about -0.73~$\mu_{B}$.
In addition, the larger $\Delta$ implies the elongation of the [Fe5-Cr] bond and lead to the reduced 
$\mu_5=2.39$~$\mu_B$ due to the tendency for a suppression of antiferromagnetism in the vicinity of Fe5. 
The increase of $\Delta$ also produces an enhancement of $\mu_6$ from 1.25~$\mu_B$ to the values 
about $1.28-1.3$~$\mu_B$, an opposite trend which occurs due to the shortening of the bond between Cr and Fe6. 

\begin{figure}[th]
\epsfxsize=5.0cm {\epsffile{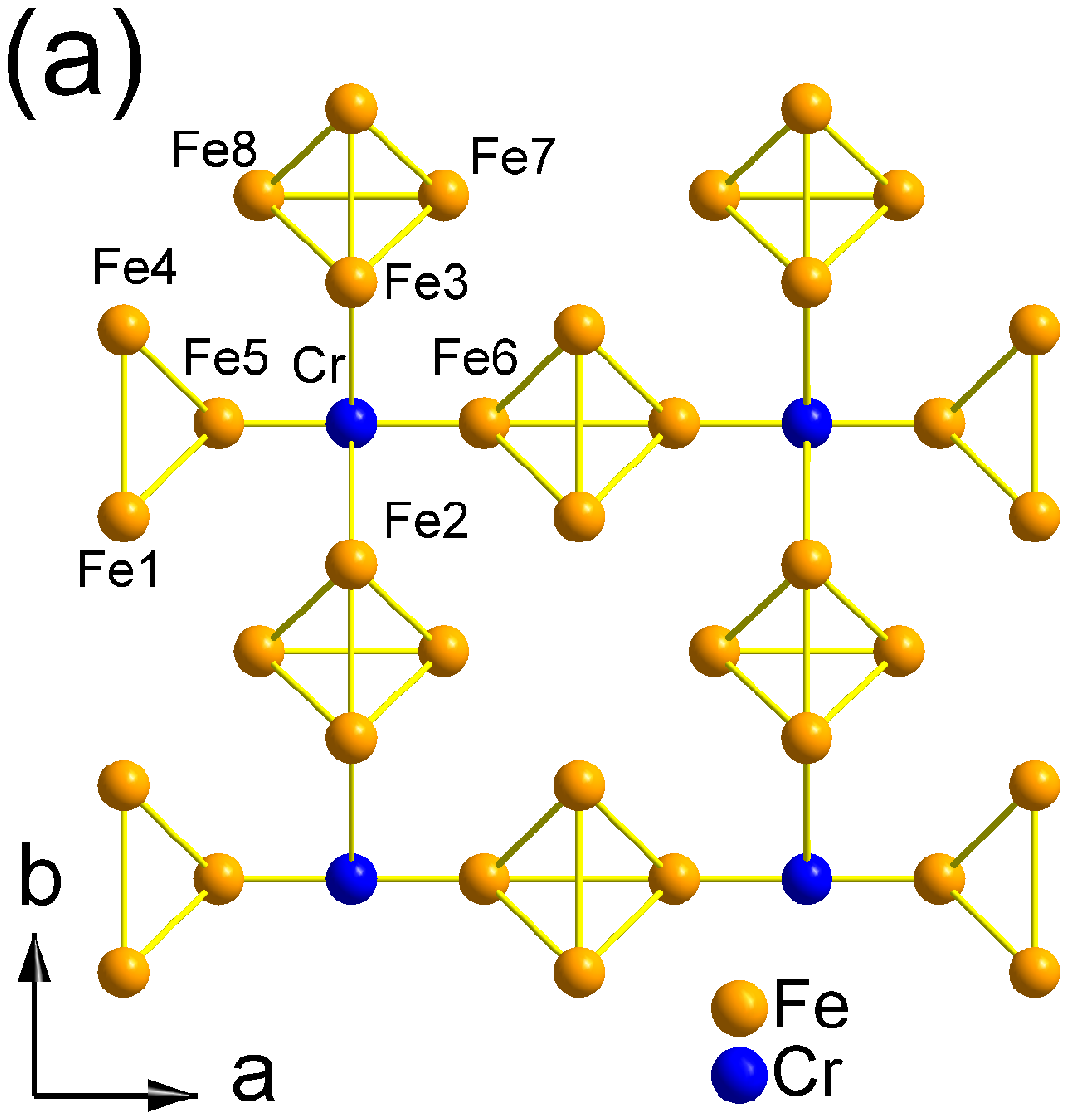}} 
\epsfxsize=5.2cm {\epsffile{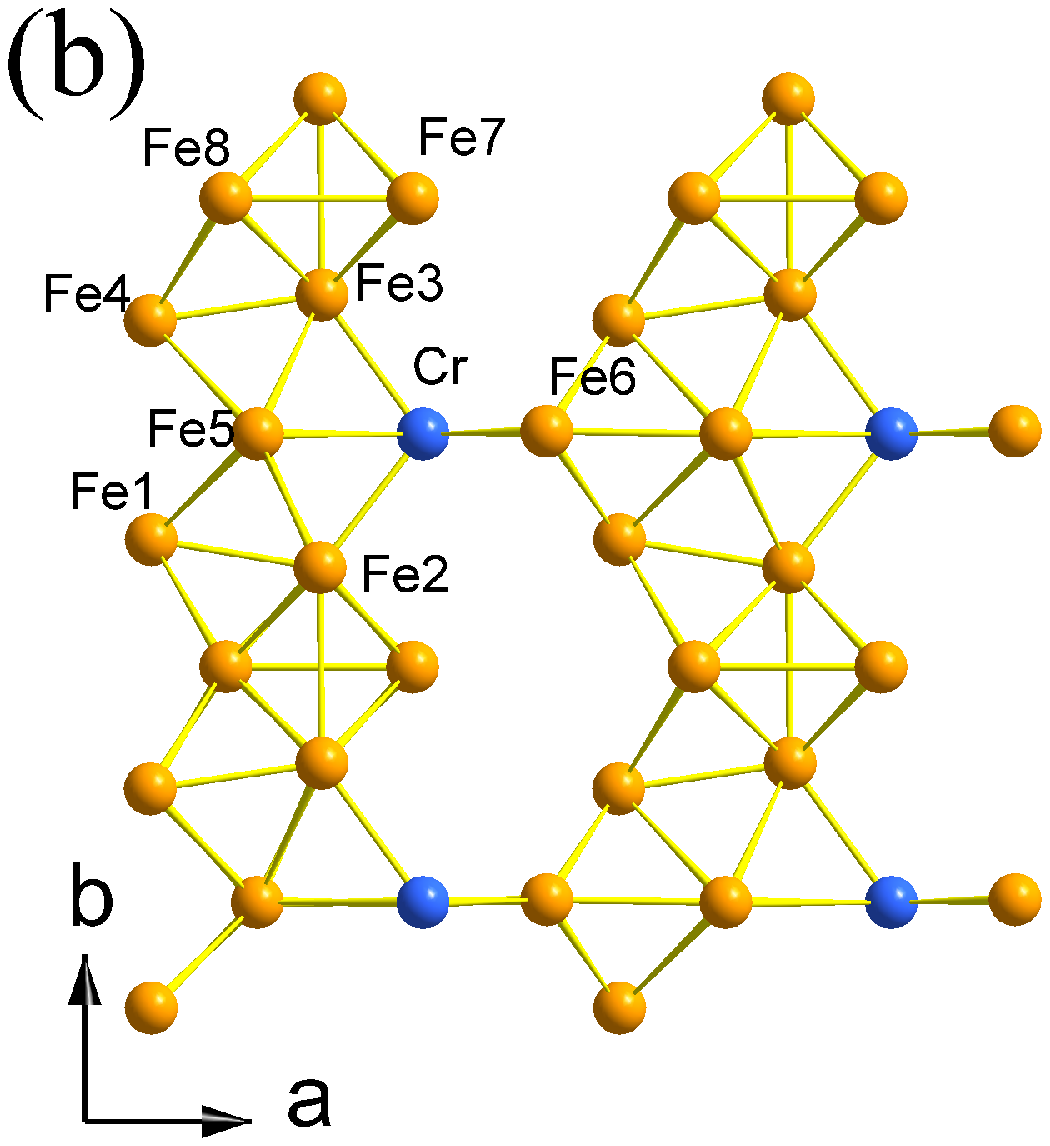}}
\epsfxsize=4.1cm {\epsffile{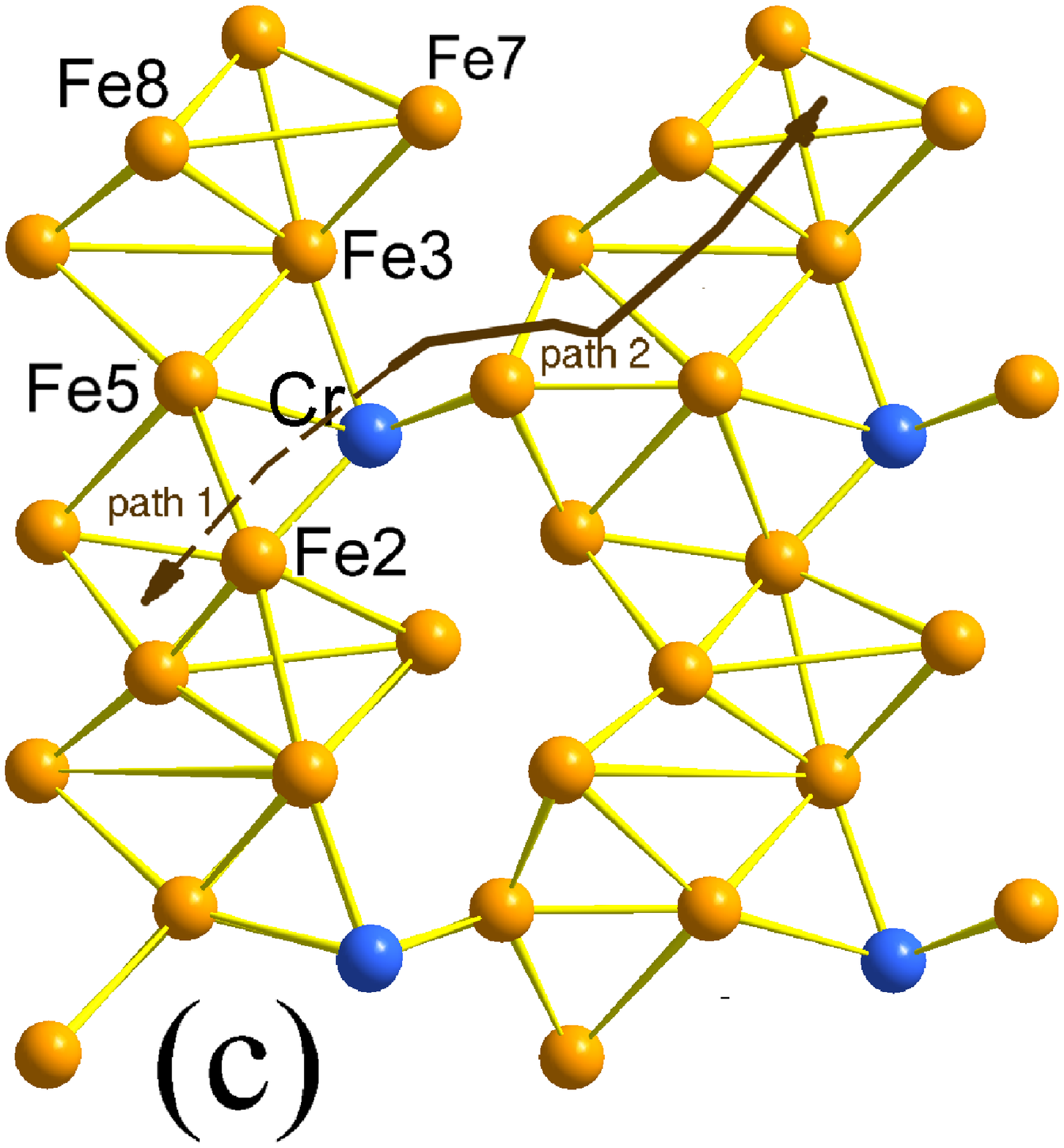}}
\caption{Relaxed structure of Fe with 12.5$\%$ of Cr: (a) LDA calculations,
(b) spin-polarized LSDA calculations in the structure with $a=b=2.86$~\AA\, and (c)
spin-polarized LSDA calculations in the structure with $a=b=3.0$~\AA. The path 1 and path 2 identify
possible pathes for diffusion through the channels formed due to atomic relaxation.
} \label{fig2}
\end{figure}

\begin{figure}[th]
\epsfxsize=8.0cm {\epsffile{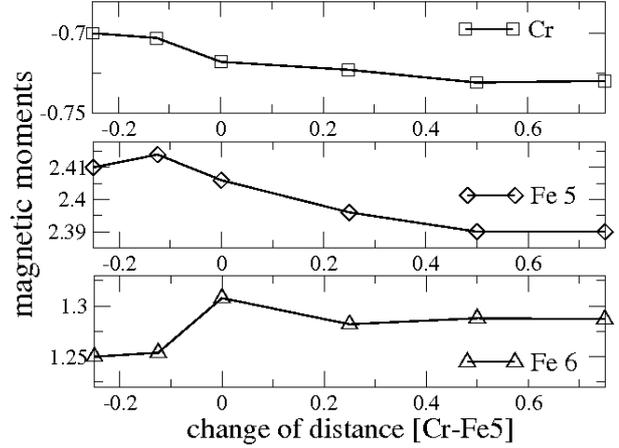}} \caption{Local magnetic moments (in $\mu_B$) of the atoms
in Fe5-Cr-Fe6 triad versus the displacement $\Delta$=[Fe5-Cr]-[Fe5-Cr]$^0$
of Cr along the (100) axis. Here [Fe5-Cr]$^0$ is the equilibrium distance between Fe6 and Cr.
} \label{fig3}
\end{figure}

In Fig.~\ref{fig3}, the $\Delta$-dependences of the atomic magnetic moments 
are highly asymmetric with respect to $\Delta$.
Consequently, the obtained magnetoelastic coupling produces an anisotropy of the 
magnetic moments and is accompanied by the loss of the inversion center due to the atomic displacements, the effect which can be 
observed in Fig.~\ref{fig2}(b) and (c). In Fig.~\ref{fig2}(c), the low-symmetry structure corresponds to the 
minimum of the total energy. 
As a conclusion, the neglect  of the magnetoelastic coupling in the electronic structure calculations does 
not allow to achieve a full optimization in this type of interstitial alloys. 

\subsection{Electronic structure}

\begin{figure}[th]
\epsfxsize=6.5cm {\epsffile{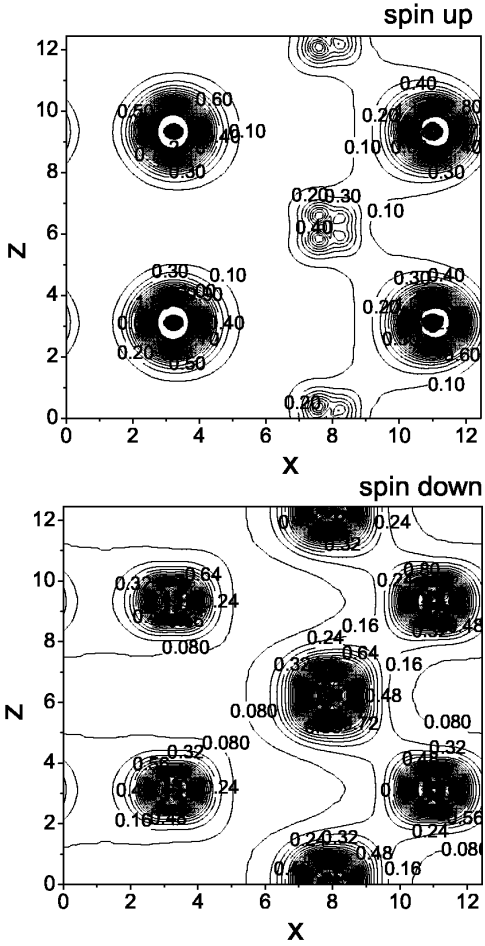}}
\caption{Contours of electron density maps in the ($x$, $z$)-plane ($y/b=0.25$, $x$ and $z$ given in~\AA) obtained by integration of 
electronic states in the energy window $E$ between $-3$~eV below the Fermi level and the Fermi level. The results obtained by the 
structural optimization using the LSDA approximation.} \label{fig4}
\end{figure}

\begin{figure}[th]
\epsfxsize=6.5cm {\epsffile{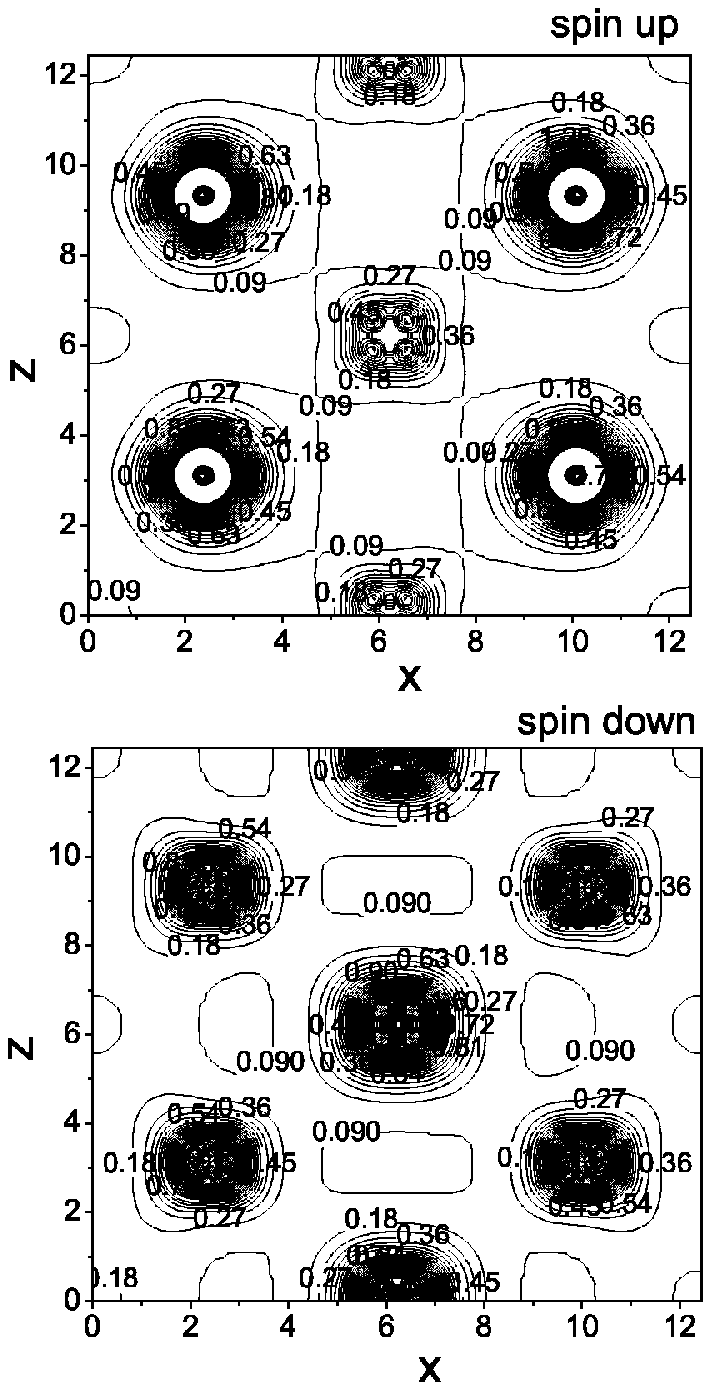}}
\caption{Contours of electron density maps in the ($x$, $z$)-plane ($y/b=0.25$, $x$ and $z$ given in~\AA) calculated by integration of 
electron states  in the energy window $E$ between $-3$~eV below the Fermi level and the Fermi level. The LSDA results obtained in 
the initially LDA-relaxed structure.} \label{fig5}
\end{figure}

Fig.~\ref{fig4} shows the $3d$ spin-polarized electronic density contours of the LSDA-optimized structure in the ($x$, $z$) 
plane.
One can see that the majority $3d$ spin-up states of Fe are highly occupied by the
electrons whereas the electron concentration of Cr spin up states is substantially lower. 
In contrast to this, the spin-down (minority) electrons are characterized by high electron occupation of Cr and lower electron
density on Fe. In Fig.~\ref{fig4}, the chain-like structures Fe-Cr in the $z$-direction are characterized by
strong hybridization between the intra-chain $3d$ spin-down orbitals of Fe and Cr. 
The last feature leads to the spatial charge redistribution 
and to higher charge densities on the bonds between spin-down Cr and Fe. In the LSDA-optimized system, 
the structural optimization produces areas with low charge density in the $y$ ($b$)-direction, where each area can be 
identified between the chains of Fe-octahedra. 
As can be seen in Fig.~\ref{fig4}, these areas are almost free of the charge and can be considered 
as channels for the migration of light atoms like H, Li or N.
Similarly to the contours in
Fig.~\ref{fig4}, the electron density of the majority Fe and Cr orbitals and on the bonds between Cr and Fe
calculated for the LDA-optimized structure (Fig.~\ref{fig5}) is substantially
lower than the charge density on the spin-down contours, although the spatial charge distribution is more
homogeneous as compared to that in Fig.~\ref{fig4}.

\begin{figure}[th]
\epsfxsize=7.0cm \centerline{\epsfclipon\epsffile{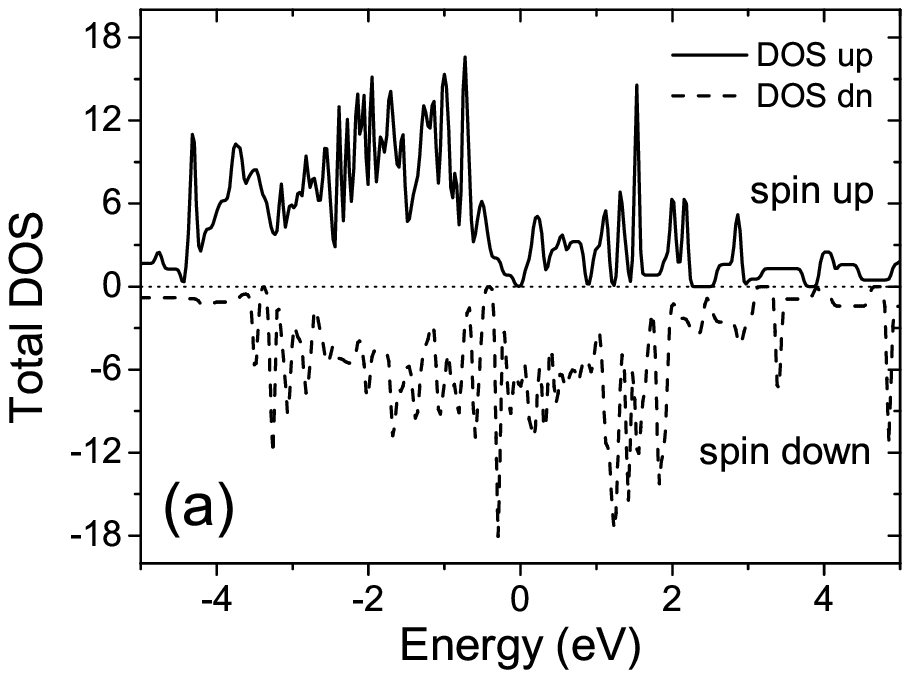}}
\epsfxsize=7.0cm \centerline{\epsfclipon\epsffile{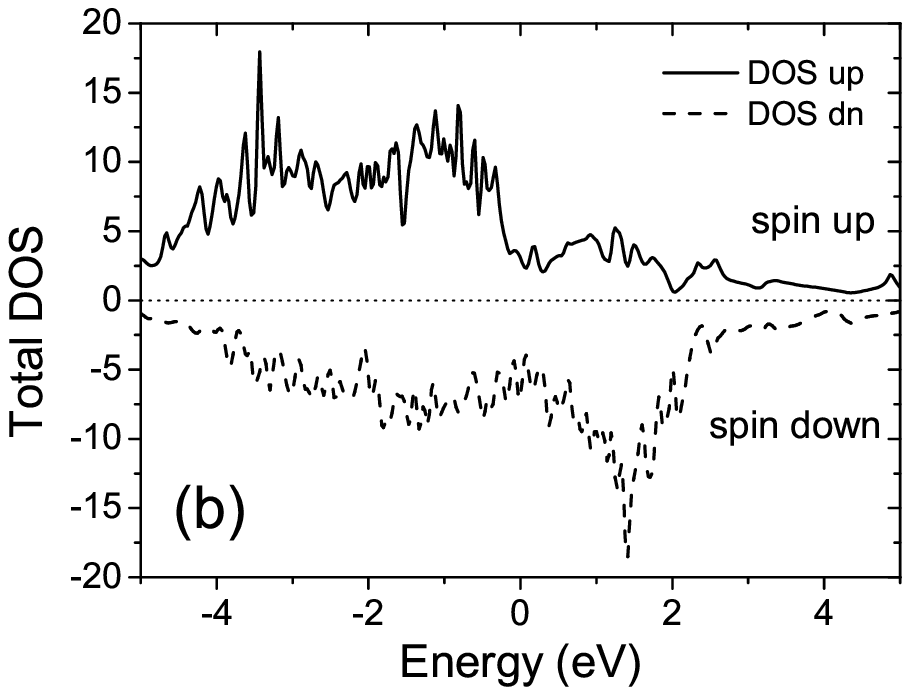}}
\caption{Total density of states (in eV$^{-1}$) for structures 
optimized using (a) LDA approach and (b) spin-polarized LSDA 
approximation. 
The Fermi level corresponds to $E=0$.} \label{fig6}
\end{figure}

For the LDA-relaxed structure, the density of states is characterized by strong suppression of the majority 
spin-up DOS at the Fermi level (Fig.~\ref{fig6}(a)), whereas the minority DOS 
at the Fermi level remains significant. Similar, although much 
stronger, suppression of majority DOS is typically observed in half-metallic systems where the electric current is 
conducted by the electrons with the same direction of spin \cite{park}. 
In contrast to the half-metallic-like features of the LDA-relaxed structure, 
the DOS of the LSDA-optimized system (Fig.~\ref{fig6}(b)) demonstrates substantial values at the Fermi level for both 
spin directions which implies an enhancement of the metallic state for the majority electrons. 

\begin{figure}[th]
\epsfxsize=8.2cm \centerline{\epsfclipon\epsffile{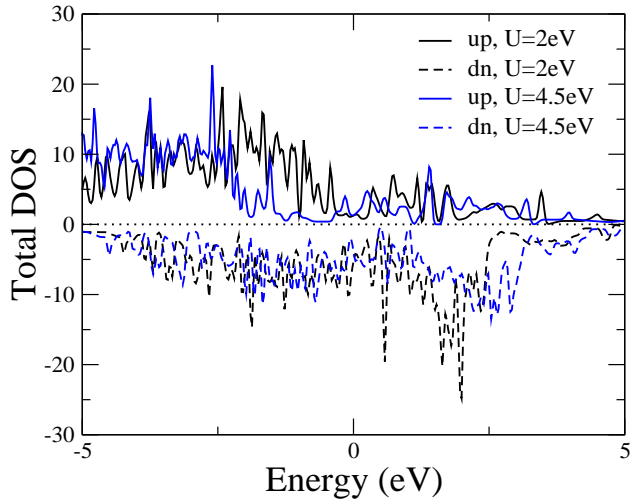}} \caption{Total densities of states  (in eV$^{-1}$) 
for the LSDA-optimized structure 
calculated by the LSDA+$U$ method with the local Coulomb corrections for the 3d-orbitals 
of Fe and Cr $U=2$~eV (black curves) and 
4~eV (blue curves). The Fermi level corresponds to $E=0$.}
\label{fig7}
\end{figure}

In transition metal oxides, the metallic state obtained in the LDA approach is usually strongly
influenced by additional account 
for the local Coulomb corrections for the $d$-electronic states \cite{anisimov,czyzyk,pavlenko3,pavlenko4}. 
In our work, the Coulomb 
corrections are incorporated within the SIC-variant of the LSDA+$U$ approximation introduced 
in Ref.~\cite{anisimov}. The results are 
presented in Fig.~\ref{fig7} for two different values of $U=2$~eV and $U=4.5$~eV estimated and employed in  
Ref.~\cite{bandyopadhyay,zhang,korotin} to account for the electron repulsion 
of 3d electrons of Fe and Cr. Fig.~\ref{fig7} shows the finite density of states at the Fermi ($E=0$) level,
although larger $U$ leads to a significant suppression of the majority DOS at $E_F$ which 
suggests a prevailing tendency towards a half-metallic behavior. 

In the LSDA-optimized structure, we find that the cell magnetic moment 
$M_{\rm LSDA}=3.84$~$\mu_B$ is larger then the magnetic moment $M_{\rm LDA}=2.88$~$\mu_B$ 
in the LDA-optimized cell.
Such an enhancement of the magnetic polarization is connected with the substantial distortions $\Delta \ve{R}_i$
in the range $0.2-0.84$~\AA\,
and can be considered as a direct evidence of significant magnetoelastic effect.
The local Coulomb corrections in the LSDA+$U$-calculations result in enhanced spin polarization.
Specifically, we obtain $M_{\rm LSDA}=4.07$~$\mu_B$ for $U=2$~eV, and $M_{\rm LSDA}=4.66$~$\mu_B$ 
for $U=4.5$~eV. It is remarkable that 
in the substitutional alloy Fe-Cr with 12.5$\%$ of Cr, the LSDA approach gives the value 3.8~$\mu_B$
for the cell magnetic moment
which is slightly lower than the magnetic moment for the considered LSDA-relaxed substitutional alloy.

The obtained high spin polarization of the considered 
interstitial alloys Fe-Cr allows us to suggest these materials as possible candidates 
for spin polarizers in the spintronic devices. 
Possible technological applications of artificially
generated interstitial alloys would be related to the thin films produced 
for the needs of modern electronic
industry. In such artificial systems, a central question is related
to the methods of implantation and positioning of Cr in centers of the edges
of cubic bcc lattice of bulk Fe. With the current state of the art, such 
structural nanoscale manipulation can be confined to the first subsurface layers of ferrite
films by the using for instance the methods of optical trapping by lasers \cite{optic}
or by atomic force microscopy \cite{afm}. To estimate the stability
of ultra-thin films of iron-chromium alloys which can be also considered as a basis for 
stainless steels, we have also extended our calculations
to the nanoscale two-monolayers-thick iron films containing one interstitial 
Cr atom per each 20-25 subsurface Fe atoms \cite{pavlenko2}.
For such type of films, we have performed the calculations of the surface formation energy
and of the electronic work functions, which were also compared to the corresponding
quantities in the films of standard substitutional Fe-Cr alloys.
In the interstitial Fe-Cr films, we find that the energy of the surface formation is 
about 1.69eV and the electronic work function amounts to 1eV, whereas for the 
substitutional Fe-Cr films we obtain 2.56eV for the surface energy 
and 0.57eV for the work function.
These results allow us to expect high stability and durability of films generated on the basis
of nanoscale-manipulated interstitial Fe-Cr alloys, as compared to the iron films
with substitutional Cr impurities.         

\section{Migration paths of N in interstitial alloys Fe$_\alpha$-Cr}

A central question related to the stability of the considered interstitial alloys
is how various atomic impurities can modify the electronic properties.
In the structurally relaxed alloys, the chains of atomic Fe-groups are separated
by 4\AA-wide atomic empty channels, which are expected to contain pathways for 
light impurity atoms like H, N or Li. To explore a possibility
of the migration of the impurities, we
consider possible migration paths of a single nitrogen atom 
in the vicinity of the atomic Fe-chains in the interstitial Fe$_\alpha$-Cr. 

In the studies of the migration paths of N impurities, we employed 
the nudged-elastic-band (NEB) method implemented in the
Quantum-Espresso (QE) Package for the DFT calculations with the use of plane-wave basis sets and
pseudopotentials \cite{qe,jonsson}. In these calculations, for the atomic
cores of Fe and Cr we employ the Perdew-Burke-Ernzerhof (PBE) norm-conserving
pseudopotentials~\cite{pbe}.
For each stage of the nitrogen transport, the NEB method
involves a relaxation of the atomic positions and of the distances between the different atoms
in the supercell  until the forces acting on the atoms reach their minima.
In these calculations, we use the plane-wave cutoff
680~eV and the energy cutoff for charge and potential given by 1360~eV. 
In the NEB-approach, the relaxation of the atomic positions along the nitrogen migration path is
performed by the minimization of the total
energy of each intermediate configuration (image).
These images correspond to different positions of N on the migration path
and they are produced by the optimization of a specially
generated object functional (action) with the consequent minimization
of the spring forces perpendicular to the path. In our calculations,
the convergence criteria for the norm of the force orthogonal to the path is 
achieved at the values below 0.05eV/\AA. As the initial atomic configuration, the supercell
Fe-Cr relaxed by the full-potential LSDA-approach~\cite{wien2k} has been considered.

Recent studies of the migration paths of single hydrogen atoms by the 
pseudopotential NEB method demonstrate a good agreement of the obtained transport
mechanisms and energy barriers with the experimental measurements \cite{pavlenko}. Similarly to 
Ref.~\cite{pavlenko}, in the present studies, each NEB-generated configuration 
has been modified by the introduction of the N atom
and the obtained in this way extended supercell has been fully structurally optimized. 

\begin{figure}[th]
\epsfxsize=5.0cm {\epsffile{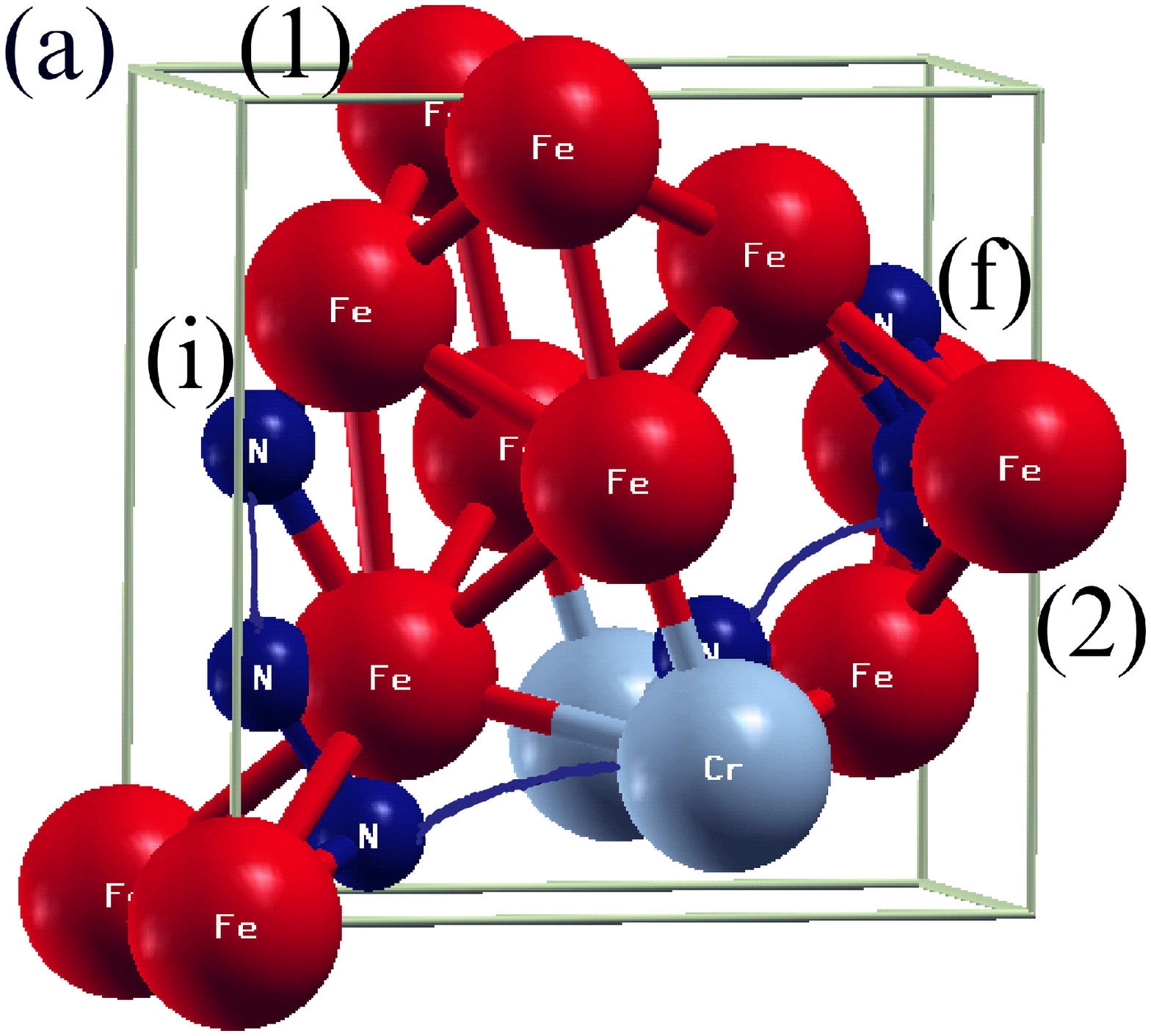}}
\epsfxsize=5.0cm {\epsffile{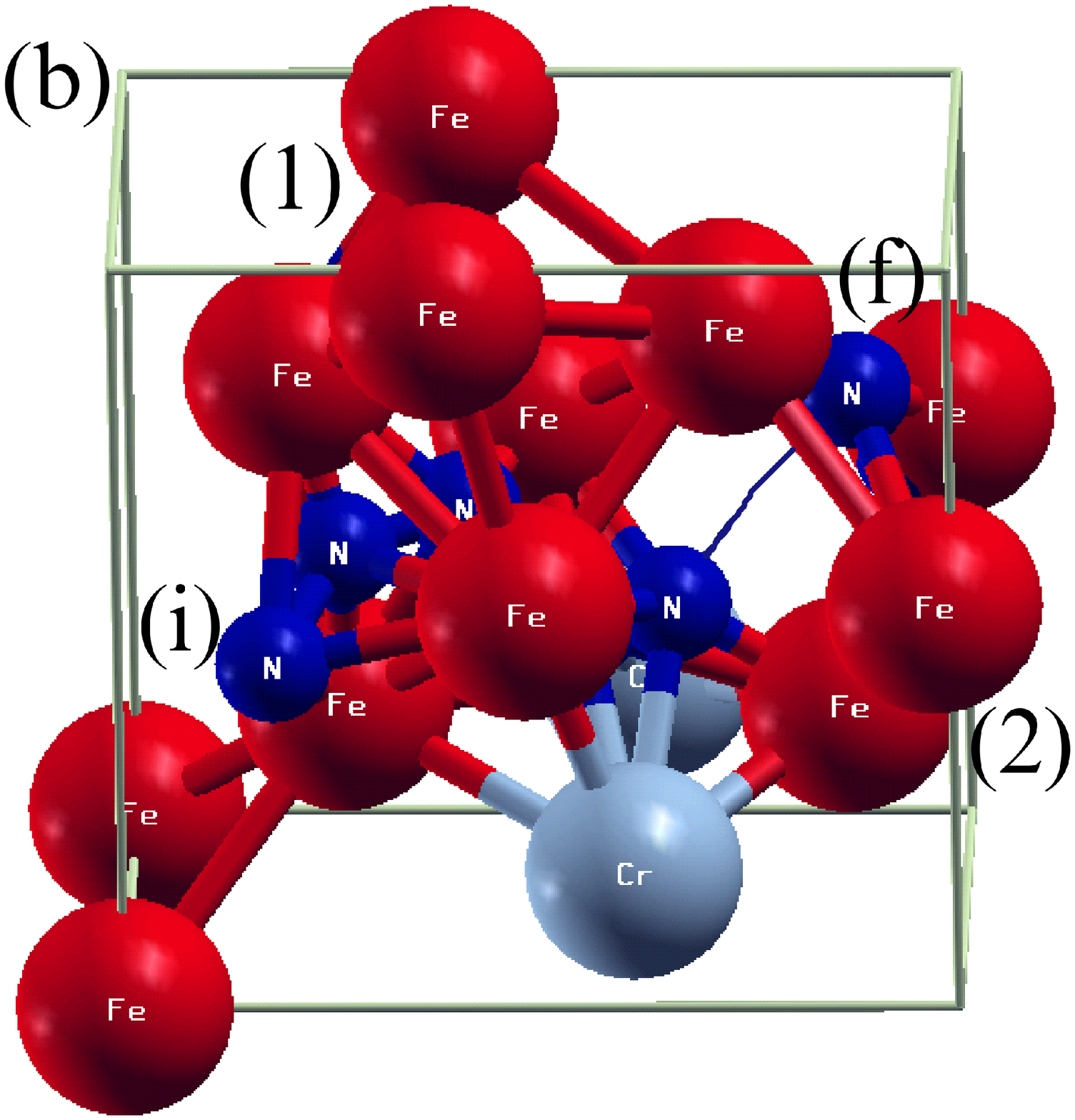}}
\caption{Two different migration paths of nitrogen through the channel
of the optimized crystal cell of Fe-Cr. The top picture (a) represents the interchain migration of N inside the
cell with the coordinate $z/c$ near $0.5$. The bottom picture (b) corresponds to the
migration of N from the cell boundary ($z=0$) along the $z$ direction with the further migration
between two neighbouring Fe-chains. The symbols (1) and (2) denote the different atomic chains; (i) and (f)
correspond to the initial and final positions of nitrogen in the migration paths.} \label{fig8}
\end{figure}

To study the migration of N, we consider two different migration paths across the atomic Fe-chains
in the Fe-Cr supercell. The first path (path (a)) describes the migration of N 
from the initial position inside the cell ($z/c=0.5$) near the chain (1) across 
the channel to the chain (2) schematically presented 
in Fig.~\ref{fig8}(a). In distinction to the path (a), the second path (b) reflects
path2 in Fig.~\ref{fig2}(c) and contains additional migration
step of N from the supercell boundary ($z\approx 0$) along the $c$-direction 
inside the supercell, with the further relocation
through the channel to the atomic Fe-chain (2) indicated in Fig.~\ref{fig8}(b). 

\begin{figure}[th]
\epsfxsize=8.0cm {\epsffile{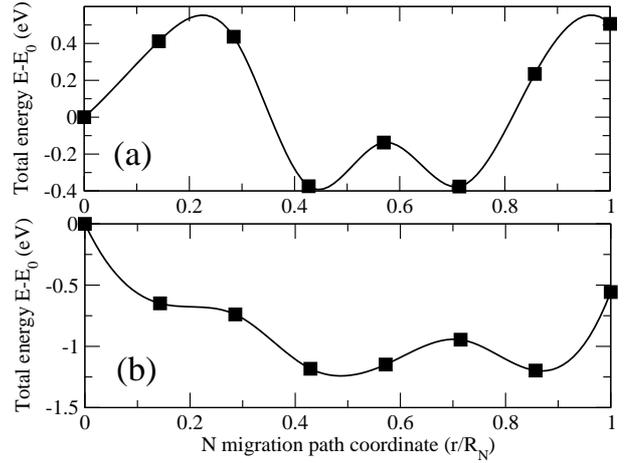}}
\caption{Total energy profiles of the system along the migration paths of N.
Here $E_0$ denotes the energy of the system in the initial position of N and $R_N$ 
is the N coordinate in the final position of the path.
} 
\label{fig9}
\end{figure}

Fig.~\ref{fig9} shows the profiles of the total energy calculated along the N migration paths (a)
and (b). The path (a) is characterized by the high energy barriers about $0.8$~eV in the path coordinate
range ($0 \le r/R_N\le 0.4$) and ($0.8 \le r/R_N\le 1$) which corresponds to the migration of N 
within the two Fe-chains (1) and (2). Here $R_N$ denotes the maximal length of the N path in the supercell
which reaches about 1~nm for the path (a).
The interchain motion inside the channel is signified by a low energy 
barrier about $0.2$~eV ($0.4 \le r/R_N\le 0.7$ in Fig.~\ref{fig9}(a)). 

In contrast to this, the energy profile
for the migration path (b) (Fig.~\ref{fig9}(b)) contains a plateau-like region at $0.1 \le r/R_N\le 0.4$. 
This miration step indicates the intracell replacement of N near the Fe-chain (1) 
along the [001]-direction demonstrated in Fig.~\ref{fig4}, with a further
relocation between the atomic chains across the atomic-empty 
channel with a low energy barrier about $0.2$~eV.

As a conclusion, we can note that the possible migration paths of the light atoms in the considered
interstitial Fe-Cr alloys contain a combination of the motion (i) within the atomic-empty channels and 
(ii) along the $c$-direction along to the Fe-contained atomic chains.

\begin{figure}[th]
\epsfxsize=8.0cm {\epsffile{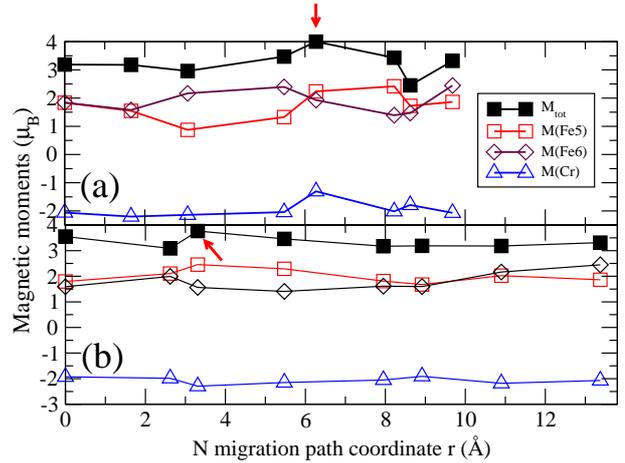}}
\caption{Cell ($M_{\rm tot}$) and atomic magnetic moments (in $\mu_B$) along the migration paths of N. 
The red arrows identify the maximal cell magnetic moments approached upon the minimization
of the distance [N-Cr] along the N migration paths.  
} \label{fig10}
\end{figure}

The question which arise due to the inclusion of N into the magnetic Fe-Cr alloy is how the N impurities 
modify the magnetic properties of the system. 
In the work of I.~Mazin\cite{mazin}, a comparison of the
degrees of spin-polarization (DSP) calculated for Fe in the static limit through the density of electronic
states, via the current densities and in the ballistic limit is presented. 
It is shown that all three definitions of the DSP give very similar
behavior for Fe due to strong hybridization of the $sp$ and $d$ states at the Fermi level.
Thus we expect that in the considered Fe alloy with relatively low concentration of Cr
it is sufficient to study the static spin polarization  in order to capture the main
properties of the alloy.
Fig.~\ref{fig10} presents the change of the cell and atomic magnetic moments at the
migration of N along the path (a) and path (b). 
Although the nitrogen is initially nonmagnetic in the bulk, it 
becomes weakly magnetic inside the Cr-Fe alloy with a small magnetic moment $-0.04$~$\mu_B$ induced by the 
magnetism of the surrounding. It is noteworthy that the cell magnetic moment is increased 
to 4~$\mu_B$ as N approaches Cr and the distance [N-Cr] becomes about 1.95~\AA. 
Such an enhancement of $M_{tot}$ is explained by the strong 
atomic distortions in the range between $0.04$~\AA (Fe7) up 
to $0.2$~\AA (Fe3) caused by the replacement of N and by the consequentmagnetoelastic effect. 
In Fig.~\ref{fig10}, the increase of the distance from N to Cr suppresses the magnetic 
moment of N and decreases the cell magnetic polarization to the typical 
values about $3.5-3.8$~$\mu_B$ obtained in LSDA-calculations for the artificial 
Fe-Cr alloys. The obtained drastic change 
of the magnetic polarization clearly demonstrates a crucial importance of the location of nonmagnetic impurities like $N$ 
for the electronic properties of alloy. As follows from our findings, a control of the location of N, for example by external 
electric field, can lead to externally tuned changes of the magnetic polarization, a feature which is of central importance 
for possible spintronic devices based on the artificial Fe-Cr alloys.

\section{Conclusion}

We have performed the DFT studies of the bulk
ferrite with $12.5\%$-concentration of monoatomic interstitial Cr
periodically located at the edges of the bcc Fe$_\alpha$ cell.
We have shown that the full atomic relaxation of the obtained
interstitial Fe-Cr stabilizes a new chain-like low-symmetry structure. In this structure, 
the monoatomic Cr at the edges of ferrite bcc cells leads to the local 
atomic distortions and results in the formation of parallel chains of Fe$_6$-ochahedra, which are 
connected by the interchain Fe-Cr bonds. The significant energy gain caused by such a structural relaxation 
approaches 6.17~eV which makes this type 
of interstitial alloy highly stable and energetically favorable with the negative formation energy approaching $-1.15$~eV. 
The novel electronic state of the system can be characterized as metallic, where the metallic properties 
is the result of strong Fe-Cr hybridization of the structurally relaxed alloy. In the investigations of the 
magnetic state of the generated relaxed 
structures, we have obtained a local antiferromagnetic order in the close proximity of Cr atoms, whereas the 
more distant Fe atoms are coupled ferromagnetically.  
We also find that the nonmagnetic impurities 
like nitrogen can substantially modify the magnetic properties of the interstitial alloy which can be considered as an 
additional manifestation of the strong magnetoelastic effect in this type of alloys. 
We propose to consider the generated 
interstitial alloys as perspective candidates 
for fabrication of novel highly durable stainless steels and for possible applications in spintronic 
and multifunctional devices.

\section{Acknowledgements}
This work has beed partially supported through the project "Models of quantum statistical 
description of catalytic processes on metallic substrates" of the Ministry of Education and Sciences
of Ukraine and the grant 0108U002091 of the National Academy of Science of Ukraine.
A grant of computer time from the Ukrainian Academic Greed is acknowledged.


\begin{thebibliography}{200}

\bibitem{steels} D.~Peckner and I.M.~Bernstein. {\it Handbook on Stainless Steels}, 
McGraw-Hill Book Co., New York, 1977.
\bibitem{mai} A.~Mai, V.A.C.~Haanappel, S.~Uhlenbruck, F.~Tietz, and D.~St\"over, Solid State Ionics, {\bf 176},
1341 (2005).
\bibitem{yokokawa} H.~Yokokawa, H.~Tu, B.~Iwanschitz, and A.~Mai, J.~Power Sources, {\bf 182},
400 (2008).
\bibitem{victora} R.H.~Victora and L.M.~Falicov, Phys.Rev.B {\bf 31},
7335 (1985).
\bibitem{paxton} A.T.~Paxton and M.W.Finnis, Phys.Rev.B {\bf 77},
024428 (2008).
\bibitem{paduani} C. Paduani and J.C. Krause, Braz. Journ. of Physics, {\bf 36},
1262 (2006).
\bibitem{davies} A.~Davies, J.A.~Stroscio, D.T.~Pierce, and R.J.~Celotta, Phys.Rev.Lett, {\bf 76}
4175 (1996).
\bibitem{klaver} T.P.C.~Klaver, P.~Olsson, and M.W.~Finnis, Phys.Rev.B {\bf 76},
214110 (2007).
\bibitem{olsson} P.~Olsson, C.~Domain, and J.~Wallenius, Phys.Rev.B {\bf 75},
014110 (2007).
\bibitem{duriagina} Z.A.Duriagina and M.I.Pashechko, Metal Science and Treatment of Metals, {\bf 4},
34 (2000).
\bibitem{afm} Y.~Sugimoto {\it et al.}, Science {\bf 322}, 413 (2008).
\bibitem{pavlenko2} N.~Pavlenko, unpublished.
\bibitem{wien2k} P.~Blaha, K.~Schwarz, G.K.H.~Madsen, D.~Kvasnicka, and J.~Luitz, {\it WIEN2K},
{\it An Augmented Plane Wave + Local Orbitals Program for Calculating Crystal Properties},
ISBN 3-9501031-1-2 (TU Wien, Austria, 2001).
\bibitem{park} J.-H.~Park {\it et al.}, Nature {\bf 392}, 794 (1998).
\bibitem{anisimov} V.I.~Anisimov, I.V.~Solovyov, M.A.~Korotin, M.T.~Czyzyk, and G.A.~Sawatzky,
Phys.Rev.~B {\bf 48}, 16929 (1993).
\bibitem{czyzyk} M.T.~Czyzyk and G.A.~Sawatzky, Phys.Rev.~B {\bf 49}, 14211 (1994).
\bibitem{pavlenko3} N.~Pavlenko, Phys.~Rev.~B {\bf 80} 075105 (2009). 
\bibitem{pavlenko4} N.~Pavlenko, I.~Elfimov, T.~Kopp, and G.A.~Sawatzky, Phys.~Rev.~B {\bf 75}, 140512(R) (2007).
\bibitem{bandyopadhyay} T.~Bandyopadhyay and D.D.~Sarma, Phys.Rev.~B {\bf 39}, 3517 (1989).
\bibitem{zhang} Ze Zhang and S.~Satpathy, Phys.Rev.~B {\bf 44}, 13319 (1991).
\bibitem{korotin} M.A.~Korotin, V.I.~Anisimov, D.I.~Khomskii, and G.A.~Sawatzky, Phys.Rev.Lett. {\bf 80}, 4305 
(1998).
\bibitem{optic} A.~Ashkin. {\it Optical trapping and manipulation of neutral particles 
using lasers}, World Scientific Pub., Singapore, 2006.
\bibitem{qe} P.~Giannozzi et al., J.~Phys.~Cond.~Matter {\bf 21}, 395502(2009).
\bibitem{jonsson} H.~Jonsson, G.~Mills, and K.W.~Jacobsen, {\it Nudged Elastic Band
Method for Finding Minimum Energy Paths of Transitions in Classical and Quantum
Dynamics in Condensed Phase Simulations} in {\it Classical and Quantum
Dynamics in Condensed Phase Simulations}, ed. by B.J.~Berne, G.~Ciccoti, and D.F.~Coker
(Singapore: World Scientific, 1998).
\bibitem{pbe} J.P.~Perdew, S.~Burke, and M.~Ernzerhof, Phys.~Rev.~Lett. {\bf 77}, 3865(1996).
\bibitem{pavlenko} N.~Pavlenko, A.~Pietraszko, A.~Pawlowski, M.~Polomska, I.V.Stasyuk, and
B.~Hilczer, Phys.~Rev.~B {\bf 84}, 064303 (2011).
\bibitem{mazin} I.~Mazin, Phys.~Rev.~Lett. {\bf 83}, 1427(1999).

\end{thebibliography}
\end{document}